\documentclass{article}
\usepackage{graphicx}
\usepackage{amsmath}

\input{tcilatex}

\begin{document}

\author{S. Manoff \\
\textit{Bulgarian Academy of Sciences}\\
\textit{\ Institute for Nuclear Research and Nuclear Energy}\\
\textit{\ Department of Theoretical Physics}\\
\textit{\ Blvd. Tzarigradsko Chaussee 72}\\
\textit{\ 1784 Sofia - Bulgaria}}
\title{Conformal derivative and conformal transports over $(\overline{L}_n,g)$
spaces}
\date{\textit{e-mail address: smanov@inrne.bas.bg}}
\maketitle

\begin{abstract}
Transports preserving the angle between two contravariant vector fields but
changing their lengths proportional to their own lengths are introduced as
''conformal'' transports and investigated over $(\overline{L}_n,g)$-spaces.
They are more general than the Fermi-Walker transports. In an analogous way
as in the case of Fermi-Walker transports a conformal covariant differential
operator and its conformal derivative are defined and considered over $(%
\overline{L}_n,g)$-spaces. Different special types of conformal transports
are determined inducing also Fermi-Walker transports for orthogonal vector
fields as special cases. Conditions under which the length of a non-null
contravariant vector field could swing as a homogeneous harmonic oscillator
are established. The results obtained regardless of any concrete field
(gravitational) theory could have direct applications in such types of
theories.

\end{abstract}

\section{Introduction}

The construction of a frame of reference for an accelerated observer by
means of vector fields preserving their lengths and the angles between them
under a Fermi-Walker transport \cite{Manoff-2}, \cite{Manoff-3}, could also
be related to the description of the motion of the axes of a gyroscope in a
space with different (not only by sign) contravariant and covariant affine
connections and metrics [$(\overline{L}_n,g)$-space]. On the other side, the
problem arises how can we describe the motion of vector fields preserving
the angles between them but, at the same time, changing the length of every
one of them proportionally to its own length. In special cases of $(%
\overline{L}_n,g)$-spaces there are different solutions of this problem
which induces the definition of a conformal transport.

\textit{Remark}. The most notions, abbreviations, and symbols in this paper
are defined in the previous papers \cite{Manoff-2}, \cite{Manoff-3}. The
reader is kindly asked to refer to one of them.

\textbf{Definition 1}. The \textit{conformal transport} is a special type of
transport (along a contravariant vector field) under which the change of the
length of a contravariant vector field is proportional to the length itself
and the angle between two contravariant vector fields does not change.

\subsection{Conformal transports in $M_n$-, $V_n$-, and $U_n$-spaces}

(a) In flat and (pseudo) Riemannian spaces without or with torsion ($M_n$-, $%
V_n$-, and $U_n$- spaces respectively, $\dim M_n=n$), [$\nabla _ug=0$ for $%
\forall u\in T(M)$] the mentioned above problem could be easily solved by
means of a conformal (angle preserving) mapping leading to the notion of 
\textit{metric conformal to a given metric} \cite{Raschewski}.

If we construct the length of $\xi $ by the use of the metric $g$ in a (for
instance) $V_n$-space as $l_\xi =\,\mid g(\xi ,\xi )\mid ^{\frac 12}$ and by
the use of the conformal to $g$ metric $\widetilde{g}=e^{2\varphi }.g$, $%
\varphi =\varphi (x^k)\in C^r(M)$, $r\geq 1$, as $\widetilde{l}_\xi =\,\mid 
\widetilde{g}(\xi ,\xi )\mid ^{\frac 12}=e^\varphi .l_\xi $, then the rate
of change $u\widetilde{l}_\xi $ of the length of $\xi $ along a
contravariant vector field $u$ leads to the relation $u\widetilde{l}_\xi
=(u\varphi ).e^\varphi .l_\xi +e^\varphi .ul_\xi =(u\varphi ).\widetilde{l}%
_\xi +e^\varphi .ul_\xi $. If $ul_\xi =0$, then $u\widetilde{l}_\xi
=(u\varphi ).\widetilde{l}_\xi $ and therefore, the change $u\widetilde{l}%
_\xi $ of the length $\widetilde{l}_\xi $ is proportional to $\widetilde{l}%
_\xi $. This is the case when $\xi $ fulfils the equation $\nabla _u\xi =0$
determining a parallel transport of $\xi $ along $u$ because of the relation
in a $M_n$-, $V_n$-, or $U_n$-space

\begin{equation}  \label{1.1}
ul_\xi =\pm \frac 1{l_\xi }.g(\nabla _u\xi ,\xi )\text{ , \thinspace
\thinspace \thinspace \thinspace }l_\xi \neq 0\text{ .}
\end{equation}

The change of the cosine of the angle $[\cos (\xi ,\eta )=(l_\xi .l_\eta
)^{-1}.g(\xi ,\eta )]$ between two contravariant vector fields $\xi $ and $%
\eta $ is described in these types of spaces by the relation 
\begin{equation*}
u[\cos (\xi ,\eta )]=\frac 1{l_\xi .l_\eta }.[g(\nabla _u\xi ,\eta )+g(\xi
,\nabla _u\eta )]- 
\end{equation*}
\begin{equation}  \label{1.2}
-[\frac 1{l_\xi }.(ul_\xi )+\frac 1{l_\eta }.(ul_\eta )].\cos (\xi ,\eta )%
\text{ , \thinspace \thinspace \thinspace \thinspace \thinspace \thinspace
\thinspace \thinspace \thinspace }l_\xi \neq 0\text{ , \thinspace \thinspace
\thinspace \thinspace }l_\eta \neq 0\text{ .}
\end{equation}

Under the conditions for parallel transport ($\nabla _u\xi =0$, $\nabla
_u\eta =0$) of $\xi $ and $\eta $, it follows that $u[\cos (\xi ,\eta )]=0$.
Since the cosine between $\xi $ and $\eta $ defined by the use of the
conformal to $g$ metric $\widetilde{g}$ as 
\begin{equation*}
\cos (\xi ,\eta )=\frac 1{\widetilde{l}_\xi .\widetilde{l}_\eta }.\widetilde{%
g}(\xi ,\eta )=\frac{e^{2\varphi }.g_{ij}.\xi ^i.\eta ^j}{e^{2\varphi }.\mid
g_{kl}.\xi ^k.\xi ^l\mid ^{\frac 12}.\mid g_{mn}.\eta ^m.\eta ^n\mid ^{\frac
12}}= 
\end{equation*}

\begin{equation}  \label{1.3}
=\frac{g_{ij}.\xi ^i.\eta ^j}{\mid g_{kl}.\xi ^k.\xi ^l\mid ^{\frac 12}.\mid
g_{mn}.\eta ^m.\eta ^n\mid ^{\frac 12}}=\frac 1{l_\xi .l_\eta }.g(\xi ,\eta )
\end{equation}

does not change by the replacement of the metric $g$ with the metric $%
\widetilde{g}$, it follows that $u[\cos (\xi ,\eta )]=0$ for $u\widetilde{l}%
_\xi =(u\varphi ).\widetilde{l}_\xi $, and $u\widetilde{l}_\eta =(u\varphi ).%
\widetilde{l}_\eta $. This means that \textit{parallel transports in a given 
}$M_n$\textit{-, }$V_n$\textit{-, or }$U_n$\textit{-space induce conformal
transports in the corresponding conformal space.}

If $ul_\xi =0$, $ul_\eta =0$ are not valid, then the proportionality of $u%
\widetilde{l}_\xi $ to $\widetilde{l}_\xi $ and of $u\widetilde{l}_\eta $ to 
$\widetilde{l}_\eta $ is violated even in these spaces. This fact induces
the problem of finding out transports different from the parallel transport (%
$\nabla _u\xi =0$, $\nabla _u\eta =0$) under which the angle between two
contravariant vector fields does not change and at the same time the rate of
change of the lengths of these vector fields is proportional to the
corresponding length.

(b) In Weyl's spaces ($\overline{W}_n$-spaces) [$\nabla _ug=\frac 1n.Q_u.g$
for $\forall u\in T(M)$, $S(dx^i,\partial _j)=f^i\,_j(x^k)$], it follows
from the general relations for $ul_\xi $ and $u[\cos (\xi ,\eta )]$ in $(%
\overline{L}_n,g)$-spaces 
\begin{equation}  \label{1.4}
ul_\xi =\pm \frac 1{2.l_\xi }.[(\nabla _ug)(\xi ,\xi )+2.g(\nabla _u\xi ,\xi
)]\text{ , \thinspace \thinspace \thinspace \thinspace \thinspace \thinspace
\thinspace \thinspace \thinspace }l_\xi \neq 0\text{ ,}
\end{equation}

\begin{equation*}
u[\cos (\xi ,\eta )]=\frac 1{l_\xi .l_\eta }.[(\nabla _ug)(\xi ,\eta
)+g(\nabla _u\xi ,\eta )+g(\xi ,\nabla _u\eta )]- 
\end{equation*}
\begin{equation}  \label{1.5}
-[\frac 1{l_\xi }.(ul_\xi )+\frac 1{l_\eta }.(ul_\eta )].\cos (\xi ,\eta )%
\text{ , \thinspace \thinspace \thinspace \thinspace \thinspace \thinspace
\thinspace \thinspace \thinspace }l_\xi \neq 0\text{ , \thinspace \thinspace
\thinspace \thinspace }l_\eta \neq 0\text{ ,}
\end{equation}

that under the conditions for parallel transport ($\nabla _u\xi =0$, $\nabla
_u\eta =0$) of $\xi $ and $\eta $ 
\begin{equation}  \label{1.6}
ul_\xi =\pm \frac 1{2.l_\xi }.\frac 1n.Q_u.g(\xi ,\xi )=\frac
1{2.n}.Q_u.l_\xi \text{ , \thinspace \thinspace \thinspace \thinspace
\thinspace \thinspace \thinspace \thinspace \thinspace \thinspace \thinspace 
}ul_\eta =\frac 1{2.n}.Q_u.l_\eta \text{ ,}
\end{equation}

\begin{equation}  \label{1.7}
u[\cos (\xi ,\eta )]=\frac 1n.Q_u.\cos (\xi ,\eta )-\frac 1n.Q_u.\cos (\xi
,\eta )=0\text{ .}
\end{equation}

Therefore, the \textit{parallel transports in }$\overline{W}_n$\textit{%
-spaces are at the same time conformal transports}.

(c) The case of $(\overline{L}_n,g)$-spaces with equal to zero trace-free
part of the covariant derivative of the metric $g$ is analogous to this of
Weyl's spaces $\overline{W}_n$.

\textit{Remark}. In $(\overline{L}_n,g)$-spaces \cite{Manoff-1} a conformal
change of a line element $ds^2=g_{\overline{i}\overline{j}%
}.dx^i.dx^j=f^k\,_i.f^l\,_j.g_{kl}.dx^i.dx^j$ could be trivially induced by
choosing $f^k\,_i=e^\varphi .g_i^k$ with $\varphi =\varphi (x^k)\in C^r(M)$
and $g_j^i$ as components of the Kronecker symbol. For this special case of
the action of a contraction operator $S$ (called in this case conformal
contraction operator $S$), the cosine of the angle between two contravariant
non-null vector fields $\xi $ and $\eta $ defined as $\cos (\xi ,\eta
)=(l_\xi .l_\eta )^{-1}.g(\xi ,\eta )$ with $l_\xi =\,\mid g(\xi ,\xi )\mid
^{\frac 12}$ and $l_\eta =\,\mid g(\eta ,\eta )\mid ^{\frac 12}$ does not
change under the action of the contraction operator $%
S:S(e^k(x^l),e_i(x^l))=f^k\,_i(x^l)\,$ [as it is the case for the
contraction operator $S=C:C(e^k(x^l),e_i(x^l))=g_i^k$] 
\begin{equation*}
\cos (\xi ,\eta )=\frac{g_{\overline{i}\overline{j}}.\xi ^i.\eta ^j}{\mid g_{%
\overline{k}\overline{l}}.\xi ^k.\xi ^l\mid ^{\frac 12}.\mid g_{\overline{m}%
\overline{n}}.\eta ^m.\eta ^n\mid ^{\frac 12}}= 
\end{equation*}
\begin{equation}  \label{1.8}
=\frac{e^{2\varphi }.g_{ij}.\xi ^i.\eta ^j}{e^{2\varphi }.\mid g_{kl}.\xi
^k.\xi ^l\mid ^{\frac 12}.\mid g_{mn}.\eta ^m.\eta ^n\mid ^{\frac 12}}=\frac{%
g_{ij}.\xi ^i.\eta ^j}{\mid g_{kl}.\xi ^k.\xi ^l\mid ^{\frac 12}.\mid
g_{mn}.\eta ^m.\eta ^n\mid ^{\frac 12}}\,\text{.}
\end{equation}

\subsection{Extended covariant differential operator over $(\overline{L}%
_n,g) $-spaces}

If one considers the transformation properties of the contravariant and
covariant affine connections over a $(\overline{L}_n,g)$-space, he can prove
that the components of every affine connection are determined up to the set
of the components of mixed tensor fields with contravariant rank 1 and
covariant rank 2, i. e. if $\nabla _{\partial _k}\partial _j=$ $\Gamma
_{jk}^i.\partial _i$ and $\Gamma _{jk}^i$ are components of a contravariant
affine connection $\Gamma $, and $\nabla _{\partial _k}dx^i=$ $P_{jk}^i.dx^j$%
, where $P_{jk}^i$ are components of a covariant affine connection $P$ in a
given (here co-ordinate) basis, then $\overline{\Gamma }\,_{jk}^i$ and $%
\overline{P}\,_{jk}^i$%
\begin{equation*}
\begin{array}{c}
\overline{\Gamma }\,_{jk}^i=\Gamma _{jk}^i-\overline{A}\,_{\,\,\,\,jk}^i%
\text{ ,\thinspace \thinspace \thinspace \thinspace \thinspace \thinspace
\thinspace \thinspace }\overline{P}\,_{jk}^i=P_{\,jk}^i-\overline{B}%
\,^i\,_{jk}\text{ , \thinspace } \\ 
\text{\thinspace \thinspace }\overline{A}=\overline{A}\,^i\,_{jk}.\partial
_i\otimes dx^j\otimes dx^k\text{ , \thinspace \thinspace \thinspace
\thinspace }\overline{B}=\overline{B}\,^i\,_{jk}.\partial _i\otimes
dx^j\otimes dx^k\text{ , \thinspace \thinspace \thinspace }\overline{A}\text{
,\thinspace \thinspace }\overline{B}\in \otimes ^1\,_2(M)\text{ ,}
\end{array}
\end{equation*}

are components (in the same basis) of a new contravariant affine connection $%
\overline{\Gamma }$ and a new covariant affine connection $\overline{P}$
respectively.

$\overline{\Gamma }$ and $\overline{P}$ correspond to a new [''extended''
with respect to $\nabla _u$, $u\in T(M)$] covariant differential operator $%
^e\nabla _u$%
\begin{equation*}
^e\nabla _{\partial _k}\partial _j=\overline{\Gamma }\,_{jk}^i.\partial _i%
\text{ ,\thinspace \thinspace \thinspace \thinspace \thinspace \thinspace
\thinspace \thinspace \thinspace \thinspace \thinspace \thinspace \thinspace
\thinspace \thinspace \thinspace \thinspace \thinspace \thinspace \thinspace
\thinspace \thinspace \thinspace \thinspace \thinspace \thinspace \thinspace
\thinspace }^e\nabla _{\partial _k}dx^i=\overline{P}\,_{jk}^i.dx^j\text{ ,} 
\end{equation*}

with the same properties as the covariant differential operator $\nabla _u$.

\textit{Remark}. On the grounds of the last relations the transformation
properties of $\overline{\Gamma }\,_{jk}^i$ and $\overline{P}\,_{jk}^i$ can
be proved in analogy to the proofs of the transformation properties of $%
\Gamma _{jk}^i$ and $P_{jk}^i$.

If we choose the tensors $\overline{A}$ and $\overline{B}$ with certain
predefined properties, then we can find $\overline{\Gamma }$ and $\overline{P%
}$ with predetermined characteristics. On the other side, $\overline{A}%
\,^i\,_{jk}$ and $\overline{B}\,^i\,_{jk}$ are related to each other on the
basis of the commutation relations of $^e\nabla _u$ and $\nabla _u$ with the
contraction operator $S$. From 
\begin{equation*}
\begin{array}{c}
\nabla _u\circ S=S\circ \nabla _u \text{ ,\thinspace \thinspace \thinspace
\thinspace \thinspace \thinspace \thinspace \thinspace \thinspace \thinspace
\thinspace \thinspace \thinspace \thinspace \thinspace \thinspace \thinspace
\thinspace \thinspace \thinspace \thinspace \thinspace \thinspace \thinspace
\thinspace \thinspace \thinspace \thinspace \thinspace }^e\nabla _u\circ
S=S\circ \,\,^e\nabla _u\text{ ,} \\ 
S(\partial _j\otimes dx^i)=S(dx^i\otimes \partial _j)=S(dx^i,\partial
_j)=f^i\,_j\text{ , \thinspace \thinspace \thinspace \thinspace \thinspace
\thinspace }f_m\,^j.f^l\,_j=g_j^i\text{ ,\thinspace \thinspace \thinspace
\thinspace \thinspace \thinspace \thinspace \thinspace }\det (f^i\,_j)\neq 0%
\text{ ,}
\end{array}
\end{equation*}

after some computations \cite{Manoff-3} we have 
\begin{equation*}
\overline{A}\,^i\,_{jk}=-\,\,\,\overline{B}\,^m\,_{lk}.f^l\,_j.f_m\,^i=-\,\,%
\,\overline{B}\,^{\underline{i}}\,_{\overline{j}k}\text{ ,\thinspace
\thinspace \thinspace \thinspace \thinspace \thinspace \thinspace \thinspace
\thinspace \thinspace \thinspace \thinspace \thinspace \thinspace \thinspace
\thinspace \thinspace \thinspace \thinspace \thinspace \thinspace \thinspace 
}\overline{B}\,^i\,_{jk}=-\,\,\,\,\overline{A}\,^l\,_{mk}.f^i\,_l.f_j\,^m=-%
\,\,\,\overline{A}\,^{\overline{i}}\,_{\underline{j}k}\text{ . } 
\end{equation*}

We can write $^e\nabla _{\partial _k}$ in the form 
\begin{equation*}
^e\nabla _{\partial _k}=\nabla _{\partial _k}-\overline{A}_{\partial _k}%
\text{ \thinspace \thinspace \thinspace \thinspace with \thinspace
\thinspace \thinspace }\overline{A}_{\partial _k}=\overline{A}%
\,^i\,_{jk}.\partial _i\otimes dx^j\text{ .} 
\end{equation*}

$^e\nabla _u$ can also be written in the form 
\begin{equation*}
^e\nabla _u=\nabla _u-\overline{A}_u\text{\thinspace \thinspace \thinspace
\thinspace \thinspace \thinspace \thinspace with\thinspace \thinspace
\thinspace \thinspace \thinspace \thinspace \thinspace }\overline{A}_u=%
\overline{A}\,^i\,_{jk}.u^k.\partial _i\otimes dx^j\text{ .} 
\end{equation*}

$\overline{A}_u$ appears as a mixed tensor field of second rank but acting
on tensor fields as a covariant differential operator because $^e\nabla _u$
is defined as covariant differential operator with the same properties as
the covariant differential operator $\nabla _u$. In fact, $\overline{A}_u$
can be defined as $\overline{A}_u=\nabla _u-\,^e\nabla _u$. If $\overline{A}%
_u$ is a given mixed tensor field, then $^e\nabla _u$ can be constructed in
a predetermined way.

In previous papers \cite{Manoff-2}, \cite{Manoff-3}, we have considered
Fermi-Walker transports over $(L_n,g)$ and $(\overline{L}_n,g)$-spaces
leading to preservation of the lengths of two contravariant vector fields
and the angle between them, when they are transported along a non-null
(non-isotropic) contravariant vector field. The investigations have been
based on a special form of the extended covariant differential operator $%
^e\nabla _u$ determined over a $(L_n,g)$- or a $(\overline{L}_n,g)$-space as 
$^e\nabla _u=\nabla _u-\overline{A}_u$, where $\nabla _u$ is the covariant
differential operator: 
\begin{equation*}
\nabla _u:v\rightarrow \nabla _uv=\overline{v}\text{ , \thinspace \thinspace
\thinspace \thinspace \thinspace \thinspace \thinspace \thinspace \thinspace
\thinspace \thinspace \thinspace }v,\,\,\overline{v}\in \otimes ^k\,_l(M)%
\text{ .} 
\end{equation*}

$u$ is a contravariant vector field, $u\in T(M)$, $v$ and $\overline{v}$ are
tensor fields with given contravariant rank $k$ and covariant rank $l$.

In accordance to its property \cite{Manoff-2} $\overline{A}_{u\,+\,\,v}=%
\overline{A}_u+\overline{A}_v$, $\overline{A}_u$ has to be linear to $u$.
The existence of covariant and contravariant metrics $g$ and $\overline{g}$
in a $(\overline{L}_n,g)$-space allows us to represent $\overline{A}_u$ in
the form $\overline{A}_u=\overline{g}(A_u)$. There are at least three
possibilities for construction of a covariant tensor field of second rank $%
A_u$ in such a way that $A_u$ is linear to $u$, i. e. we can determine $A_u$
as

A. $A_u=C(u)=A(u)=A_{ij\overline{k}}.u^k.dx^i\otimes dx^j$ .

B. $A_u=C(u)=\nabla _uB=B_{ij;k}.u^k.dx^i\otimes dx^j$ .

C. $A_u=C(u)=A(u)+\nabla _uB=(A_{ij\overline{k}}+B_{ij;k}).u^k.dx^i\otimes
dx^j$.

These three possibilities for definition of $A_u$ lead to three types ($A$, $%
B$, and $C$ respectively) of the extended covariant differential operator $%
^e\nabla _u=\nabla _u-\overline{g}(A_u)$. It can obey additional conditions
determining the structure of the mixed tensor field $\overline{A}_u=%
\overline{g}(A_u)$. One can impose given conditions on $^e\nabla _u$ leading
to predetermined properties of $\overline{A}_u$ and vice versa: one can
impose conditions on the tensor field $\overline{A}_u$ leading to
predetermined properties of $^e\nabla _u$. The method of finding out the
conditions for Fermi-Walker transports over $(\overline{L}_n,g)$-spaces
allow us to consider other types of transports with important properties for
describing the motion of physical systems over such type of spaces.

As it was mentioned above an interesting problem is the finding out of
transports along a contravariant vector field under which the change of the
length of a contravariant vector field is proportional to the length itself
and, at the same time, the angle between two contravariant vector fields
does not change. These preconditions for a transport lead to its definition
and its applications as a conformal transport over $(\overline{L}_n,g)$%
-spaces.

In Sec. 2 conformal transports over $(\overline{L}_n,g)$-spaces are
determined and considered with respect to their structure. A conformal
covariant differential operator and its corresponding conformal derivative
are introduced. In Sec. 3 Fermi-Walker transports along a contravariant
vector field for orthogonal to it contravariant vector fields are found on
the basis of the structure of conformal transports. In Sec. 4 conformal
transports of null vector fields are discussed. In Sec. 5 the length of a
contravariant vector field as homogeneous harmonic oscillator with given
frequency is considered on the grounds of a conformal transport. Sec. 6
comprises some concluding remarks.

\textit{Remark}. All formulas written in index-free form are identical and
valid in their form (but not in their contents) for $(L_n,g)$- and $(%
\overline{L}_n,g)$-spaces. The difference between them appears only if they
are written in a given (co-ordinate or non-co-ordinate) basis.

\section{Conformal transports over $(\overline{L}_n,g)$-spaces}

Let us now take a closer look at the notion of conformal transport over $(%
\overline{L}_n,g)$-spaces.

\textit{Remark}. The following below conditions are introduced on the
analogy of the case of Fermi-Walker transports.

The main assumption related to the notion of conformal transport and leading
to a definition of an external covariant differential operator $^e\nabla
_u=\,^c\nabla _u$ (called conformal covariant differential operator) is that
the parallel transports $^c\nabla _u\xi =0$ and $^c\nabla _u\eta =0$ of two
contravariant non-null vector fields $\xi $ and $\eta $ respectively induce
proportional to the lengths $l_\xi =\,\mid g(\xi ,\xi )\mid ^{\frac 12}$ and 
$l_\eta =\,\mid g(\eta ,\eta )\mid ^{\frac 12}$ changes of $l_\xi $ and $%
l_\eta $ along a contravariant vector field $u$ as well as preservation of
the angle between $\xi $ and $\eta $ with respect to a special transport of
the covariant differential operator $\nabla _u$. The rate of change of the
length $l_\xi $ of a non--null contravariant vector field $\xi $ and the
rate of change of the cosine of the angle between two non-null contravariant
vector fields $\xi $ and $\eta $ over a $(\overline{L}_n,g)$-space can be
found in the forms (\ref{1.4}) and (\ref{1.5}). If $^c\nabla _u\xi =\nabla
_u\xi -\overline{A}_u\xi =0$ and $^c\nabla _u\eta =\nabla _u\eta -\overline{A%
}_u\eta =0$, then the conditions have to be fulfilled 
\begin{equation}  \label{2.1}
ul_\xi =\pm \frac 1{2.l_\xi }.[(\nabla _ug)(\xi ,\xi )+2.g(\overline{A}_u\xi
,\xi )]=f(u).l_\xi \text{ , \thinspace \thinspace \thinspace \thinspace
\thinspace \thinspace \thinspace }l_\xi \neq 0\text{ , \thinspace \thinspace
\thinspace \thinspace \thinspace }f(u)\in C^r(M)\text{ ,}
\end{equation}
\begin{equation*}
u[\cos (\xi ,\eta )]=\frac 1{l_\xi .l_\eta }.[(\nabla _ug)(\xi ,\eta )+g(%
\overline{A}_u\xi ,\eta )+g(\xi ,\overline{A}_u\eta )]- 
\end{equation*}
\begin{equation}  \label{2.2}
-[\frac 1{l_\xi }.(ul_\xi )+\frac 1{l_\eta }.(ul_\eta )].\cos (\xi ,\eta )=0%
\text{ .}
\end{equation}

From the first condition, under the assumption $\overline{A}_u=\overline{g}%
(A_u)$ with $A_u=C(u)=C_{ij}(u).dx^i\otimes dx^j$, we obtain 
\begin{equation}  \label{2.3}
ul_\xi =\pm \frac 1{2.l_\xi }.[(\nabla _ug)(\xi ,\xi )+2.g(\overline{g}%
(C(u))(\xi ),\xi )]=f(u).l_\xi \text{ ,}
\end{equation}

where \cite{Manoff-2} 
\begin{equation}  \label{2.4}
\begin{array}{c}
g( \overline{g}(C(u))(\xi ),\xi )=[_sC(u)](\xi ,\xi )=\,_sC_{jk}(u).\xi ^{%
\overline{j}}.\xi ^{\overline{k}}\text{ ,} \\ 
_sC(u)=C_{(jk)}(u).dx^j.dx^k \text{ , \thinspace \thinspace \thinspace }%
C_{(jk)}(u)=\frac 12.[C_{jk}(u)+C_{kj}(u)] \\ 
\text{\thinspace \thinspace }dx^j.dx^k=\frac 12(dx^j\otimes dx^k+dx^k\otimes
dx^j)\text{ .}
\end{array}
\end{equation}

On the other hand, $(\nabla _ug)(\xi ,\xi )=g_{jk;m}.u^m.\xi ^{\overline{j}%
}.\xi ^{\overline{k}}$. Therefore, the condition $ul_\xi =f(u).l_\xi $ for $%
\forall \xi \in T(M)$ leads to the relations 
\begin{equation*}
ul_\xi =\pm \frac 1{2.l_\xi }.\{(\nabla _ug)(\xi ,\xi )+2.[_sC(u)](\xi ,\xi
)\}=f(u).l_\xi \text{ , \thinspace \thinspace \thinspace \thinspace
\thinspace \thinspace \thinspace }l_\xi \neq 0\text{ , \thinspace \thinspace
\thinspace \thinspace \thinspace }f(u)\in C^r(M)\text{ ,} 
\end{equation*}
\begin{equation}  \label{2.5}
\begin{array}{c}
(\nabla _ug)(\xi ,\xi )+2.[_sC(u)](\xi ,\xi )=\pm 2.f(u).l_\xi
^2=2.f(u).g(\xi ,\xi ) \text{ for }\forall \xi \in T(M)\text{ ,} \\ 
\lbrack \nabla _ug+2._sC(u)-2.f(u).g](\xi ,\xi )=0 \text{ for }\forall \xi
\in T(M)\text{ ,} \\ 
\nabla _ug=-2._sC(u)+2.f(u).g:\,_sC(u)=-\frac 12.\nabla _ug+f(u).g\text{ .}
\end{array}
\end{equation}

Since $C(u)$ and respectively $_sC(u)$ have to be linear to $u$, $f(u)$
should have the form $f(u)=f_{\overline{k}}.u^k$,\thinspace \thinspace
\thinspace $f\in T^{*}(M)$. Therefore, 
\begin{equation}  \label{2.6}
_sC(u)=-\,\frac 12.\nabla _ug+f(u).g\text{ , \thinspace \thinspace
\thinspace \thinspace \thinspace \thinspace }u\in T(M)\text{ , \thinspace
\thinspace \thinspace \thinspace }f\in T^{*}(M)\text{ , for }\forall \xi \in
T(M)\text{ .}
\end{equation}

By the use of the condition $ul_\xi =f(u).l_\xi $ for $\forall \xi \in T(M)$
and $l_\xi \neq 0$, we have found the explicit form of the symmetric part $%
_sC(u)$ of $C(u)=\,_sC(u)+\,_aC(u)$, where 
\begin{equation}  \label{2.7}
\begin{array}{c}
_aC(u)=C_{[jk]}(u).dx^j\wedge dx^k \text{ , \thinspace \thinspace }%
C_{[jk]}(u)=\frac 12.[C_{jk}(u)-C_{kj}(u)]\text{ ,} \\ 
dx^j\wedge dx^k=\frac 12(dx^j\otimes dx^k-dx^k\otimes dx^j)\text{ .}
\end{array}
\end{equation}

If we now assume the validity of the first condition $ul_\xi =f(u).l_\xi $
[fulfilled for $_sC(u)=-\,\frac 12.\nabla _ug+f(u).g$], then from the
expression for $u[\cos (\xi ,\eta )]$ we obtain the relations: 
\begin{equation*}
u[\cos (\xi ,\eta )]=\frac 1{l_\xi .l_\eta }.[(\nabla _ug)(\xi ,\eta )+g(%
\overline{A}_u\xi ,\eta )+g(\xi ,\overline{A}_u\eta )]- 
\end{equation*}
\begin{equation}  \label{2.8}
-[\frac 1{l_\xi }.f(u).l_\xi +\frac 1{l_\eta }.f(u).l_\eta ].\cos (\xi ,\eta
)\text{ , \thinspace \thinspace \thinspace for }\forall \xi ,\eta \in T(M)%
\text{ ,}
\end{equation}
\begin{equation}  \label{2.9}
\begin{array}{c}
g( \overline{A}_u\xi ,\eta )=g(\overline{g}(C(u))(\xi ),\eta )=g_{\overline{i%
}\overline{j}}.g^{il}.C_{\overline{l}\overline{k}}(u).\xi ^k.\eta
^j=g_j^l.C_{\overline{l}\overline{k}}(u).\xi ^k.\eta ^j= \\ 
=\,C_{\overline{j}\overline{k}}(u).\eta ^j.\xi ^k=\,C_{jk}(u).\eta ^{%
\overline{j}}.\xi ^{\overline{k}}=[C(u)](\eta ,\xi )\text{ ,} \\ 
g( \overline{A}_u\eta ,\xi )=g(\overline{g}(C(u))(\eta ),\xi )=g_{\overline{i%
}\overline{j}}.g^{il}.C_{\overline{l}\overline{k}}(u).\eta ^k.\xi
^j=g_j^l.C_{\overline{l}\overline{k}}(u).\eta ^k.\xi ^j= \\ 
=\,C_{\overline{j}\overline{k}}(u).\xi ^j.\eta ^k=\,C_{jk}(u).\xi ^{%
\overline{j}}.\eta ^{\overline{k}}=[C(u)](\xi ,\eta )\text{ ,} \\ 
(\nabla _ug)(\xi ,\eta )+g( \overline{A}_u\xi ,\eta )+g(\xi ,\overline{A}%
_u\eta )=(\nabla _ug)(\xi ,\eta )+[_sC(u)](\eta ,\xi )+[_aC(u)](\eta ,\xi )+
\\ 
+[_sC(u)](\xi ,\eta )+[_aC(u)](\xi ,\eta )\text{ .}
\end{array}
\end{equation}

Since $[_sC(u)](\eta ,\xi )=[_sC(u)](\xi ,\eta )=-\,\frac 12.(\nabla
_ug)(\xi ,\eta )+f(u).g(\xi ,\eta )$ and $[_aC(u)](\eta ,\xi )=-[_aC(u)](\xi
,\eta )$, it follows for $u[\cos (\xi ,\eta )]$%
\begin{equation*}
u[\cos (\xi ,\eta )]=\frac 1{l_\xi .l_\eta }.\{(\nabla _ug)(\xi ,\eta
)+2.[_sC(u)](\xi ,\eta )\}-2.f(u).\cos (\xi ,\eta )= 
\end{equation*}
\begin{equation*}
=\frac 1{l_\xi .l_\eta }.\{(\nabla _ug)(\xi ,\eta )+2.f(u).g(\xi ,\eta
)-(\nabla _ug)(\xi ,\eta )\}-2.f(u).\cos (\xi ,\eta )= 
\end{equation*}
\begin{equation}  \label{2.10}
=\frac 1{l_\xi .l_\eta }.2.f(u).l_\xi .l_\eta .\cos (\xi ,\eta )-2.f(u).\cos
(\xi ,\eta )=0\text{ , \thinspace \thinspace \thinspace \thinspace }g(\xi
,\eta )=l_\xi .l_\eta .\cos (\xi ,\eta )\text{ .}
\end{equation}

Therefore, $C(u)$ will have the explicit form 
\begin{equation}  \label{2.11}
C(u)=\,_aC(u)-\frac 12.\nabla _ug+f(u).g\text{ .}
\end{equation}

Now, we can define the notion of conformal covariant differential operator.

\textbf{Definition 2}. A \textit{conformal covariant differential operator} $%
^c\nabla _u$ in a $(\overline{L}_n,g)$-space. An extended covariant operator 
$^e\nabla _u$ with the structure 
\begin{equation*}
^e\nabla _u=\,^c\nabla _u=\nabla _u-\overline{g}(C(u))\text{ ,} 
\end{equation*}

where 
\begin{equation*}
C(u)=\,_aC(u)-\frac 12.\nabla _ug+f(u).g\text{ , \thinspace \thinspace } 
\end{equation*}
\begin{equation*}
\text{\thinspace }C(u)\in \otimes _2(M)\text{, \thinspace \thinspace
\thinspace \thinspace \thinspace \thinspace \thinspace }_aC(u)\in \Lambda
^2(M)\text{ ,\thinspace \thinspace \thinspace \thinspace \thinspace
\thinspace }u\in T(M)\text{ ,\thinspace \thinspace \thinspace \thinspace
\thinspace \thinspace }f\in T^{*}(M)\text{ , \thinspace \thinspace
\thinspace }g\otimes _{s2}(M)\text{ .} 
\end{equation*}

is called conformal covariant differential operator. It is denoted as $%
^c\nabla _u$.

\textit{Remark}. For $f(u)=0:\,^{\,c}\nabla _u=\,^F\nabla _u$, i. e. for $%
f(u)=0 $ the conformal covariant differential operator $^c\nabla _u$ is
identical with the Fermi covariant differential operator $^F\nabla _u$.

On the analogy of the case of $^F\nabla _u$ we can have three types $%
(A,\,B,\,$and $C)$ of $^c\nabla _u$:

\small%
%
\textbf{Table 1}. Types of conformal covariant differential operator $%
^c\nabla _u$

\begin{center}
\fbox{$
\begin{array}{ll}
\underline{\text{Type of }^c\nabla _u} & \underline{\text{Form of }C(u)} \\ 
A & C(u)=A(u)=A_{ij \overline{k}}.u^k.dx^i\otimes dx^j \\ 
B & C(u)=\nabla _uB=B_{ij;k}.u^k.dx^i\otimes dx^j \\ 
C & C(u)=A(u)+\nabla _uB=(A_{ij\overline{k}}+B_{ij;k}).u^k.dx^i\otimes dx^j
\end{array}
$}
\end{center}

\normalsize%
%

The first two types $A$ and $B$ are special cases of type $C$.

From the explicit form of $C(u)=\,_aC(u)-\frac 12\nabla _ug+f(u).g$ we can
choose $_aC(u)=\,_aA(u)$ and $_sC(u)=\,_sA(u)+\nabla _uB$ with $%
_sA(u)=f(u).g $ and $B=-\frac 12.\nabla _ug$.

A type of general ansatz (without too big loss of generality) for $%
A(u)=\,_aA(u)+\,_sA(u)=\,_aA(u)+f(u).g$ (keeping in mind its linearity to $u$
and the form of $\overline{A}_u=u^k.\overline{A}_{\partial _k}$) has the
form \cite{Manoff-2} 
\begin{equation}  \label{2.12}
A_{ij}(u)=p_{\overline{k}}.u^k.^F\omega _{ij}+f_{\overline{k}}.u^k.g_{ij}%
\text{ ,}
\end{equation}

or 
\begin{equation}  \label{2.13}
A(u)=p(u).^F\omega +f(u).g\text{ ,}
\end{equation}

where 
\begin{equation*}
^F\omega _{ij}=-\,^F\omega _{ji}\text{ , \thinspace \thinspace \thinspace }%
_aA(u)=\frac 12.[A_{ij}(u)-A_{ji}(u)].dx^i\wedge dx^j\text{ ,} 
\end{equation*}

$p,f\in T^{*}(M)$ are arbitrary given covariant vector fields, $p(u)=p_{%
\overline{k}}.u^k=S(p,u)$,\thinspace \thinspace \thinspace $f(u)=f_{%
\overline{k}}.u^k=S(f,u)$, $^F\omega =\,^F\omega _{ij}.dx^i\wedge dx^j$ is
an arbitrary given covariant antisymmetric tensor field of second rank.
Therefore, $\overline{g}(A(u))=\overline{g}(p(u).^F\omega )+\overline{g}%
(f(u).g)=p(u).\overline{g}(^F\omega )+f(u).\overline{g}(g)$, $\overline{g}%
(g)=g^{i\overline{j}}.g_{jk}.\partial _i\otimes dx^k$. If we express $p$ and 
$f$ by the use of their corresponding with respect to the metric $g$
contravariant vector fields $b=\overline{g}(p):g(b)=p$, and $q=\overline{g}%
(f):g(q)=f$ respectively, then $\overline{A}_u$ will obtain the form 
\begin{equation}  \label{2.14}
\overline{A}_u=\overline{g}(C(u))=g(b,u).\overline{g}(^F\omega )+g(q,u).%
\overline{g}(g)-\frac 12.\overline{g}(\nabla _ug)\text{ , \thinspace
\thinspace \thinspace \thinspace \thinspace \thinspace }b,q\in T(M)\text{ .}
\end{equation}

The components $\overline{A}\,^i\,_{jk}$ of $\overline{A}_{\partial _k}$ in
a co-ordinate basis have the form 
\begin{equation}  \label{2.14a}
\overline{A}\,^i\,_{jk}=g^{im}.^F\omega _{\overline{m}j}.g_{\overline{k}%
l}.b^l+g_{\overline{k}\overline{l}}.q^l.g^{im}.g_{\overline{m}j}-\frac
12.g^{i\overline{m}}.g_{mj;k}\text{ .}
\end{equation}

Respectively, the components $\overline{B}\,^i\,_{jk}$ of $\overline{B}%
_{\partial _k}$ in a co-ordinate basis have the form 
\begin{equation}  \label{2.14b}
\overline{B}\,^i\,_{jk}=-\overline{A}\,^{\overline{i}}\,_{\underline{j}k}=g^{%
\overline{i}m}.^F\omega _{\overline{m}\underline{j}}.g_{\overline{k}%
\overline{l}}.b^l+g_{\overline{k}\overline{l}}.q^l.g^{\overline{i}m}.g_{%
\overline{m}\underline{j}}-\frac 12.g^{\overline{i}\overline{m}}.g_{m%
\overline{j};k}\text{ .}
\end{equation}

We have now the free choice of the contravariant vector fields $b$ and $q$,
which could depend on the physical problem to be considered. For the
determination of Fermi-Walker transports the vector field $b$ has been
chosen as $b=\frac 1e.u$ with $e=g(u,u)\neq 0$. There are other
possibilities for the choice of $b$ and $q$.

The conformal covariant differential operator will now have the form 
\begin{equation}  \label{2.15}
^c\nabla _u=\nabla _u-\overline{A}_u=\nabla _u-[g(b,u).\overline{g}(^F\omega
)+g(q,u).\overline{g}(g)-\frac 12.\overline{g}(\nabla _ug)]\text{ .}
\end{equation}

The result $^c\nabla _uv$ of the action of a conformal covariant
differential operator $^c\nabla _u$ (of type $A$, $B$, and $C$) on a tensor
field $v\in \otimes ^k\,_l(M)$ is called \textit{conformal derivative} of
type $A$, $B$, and $C$ respectively of the tensor field $v$.

\section{Fermi-Walker transports for orthogonal to $u$ vector fields}

The free choice of the vector fields $b$ and $q$ allows us to determine
another type of Fermi-Walker transport for orthogonal to $u$ vector fields
than the defined in \cite{Manoff-2}.

(a) If we chose $b=\frac 1e.u$ and $q=\xi $ in the expression for $^c\nabla
_u\xi $ we will have 
\begin{equation}  \label{2.16}
^c\nabla _u\xi =\nabla _u\xi -[\overline{g}(^F\omega )(\xi )+l.\xi -\frac 12.%
\overline{g}(\nabla _ug)(\xi )]\text{ ,}
\end{equation}

where $l=g(\xi ,u)$, $\overline{g}(g)(\xi )=\overline{g}[g(\xi )]=\xi $.
Then $^c\nabla _u\xi =0$ is equivalent to a conformal transport in the form 
\begin{equation}  \label{2.17}
\nabla _u\xi =[\overline{g}(^F\omega )(\xi )+l.\xi -\frac 12.\overline{g}%
(\nabla _ug)(\xi )]\text{ .}
\end{equation}

It is obvious that if the vector field $\xi $ is orthogonal to $u$ $%
[l=g(u,\xi )=0]$, then $\nabla _u\xi =[\overline{g}(^F\omega )(\xi )-\frac
12.\overline{g}(\nabla _ug)(\xi )]=\overline{g}[^F\omega (\xi )]-\frac 12.%
\overline{g}[(\nabla _ug)(\xi )]$ is a generalized Fermi-Walker transport of
type $C$ along a non-null vector field $u$.

\textit{Remark}. If $q=\xi $ (or $b=\xi $, or $b=q=\xi $) in the expression
for $^c\nabla _u\xi $, then $^c\nabla _u\xi $ is not more a linear transport
with respect to the contravariant vector field $\xi $.

(b) If we chose $b=q=\xi $ in the expression for $^c\nabla _u\xi $, it
follows that 
\begin{equation}  \label{2.18}
^c\nabla _u\xi =\nabla _u\xi -\{l.[\overline{g}(^F\omega )(\xi )+\xi ]-\frac
12.\overline{g}(\nabla _ug)(\xi )\}\text{ .}
\end{equation}

Then $^c\nabla _u\xi =0$ is equivalent to a conformal transport in the form

\begin{equation}  \label{2.19}
\nabla _u\xi =l.[\overline{g}(^F\omega )(\xi )+\xi ]-\frac 12.\overline{g}%
(\nabla _ug)(\xi )\text{ .}
\end{equation}

For an orthogonal to $u$ vector field $\xi $ $[l=g(\xi ,u)=0]$ we obtain a
Fermi-Walker transport of type $B$ for the vector field $\xi $%
\begin{equation}  \label{2.20}
\nabla _u\xi =-\frac 12.\overline{g}(\nabla _ug)(\xi )=\overline{g}[(\nabla
_uB)(\xi )]
\end{equation}

with $B=-\frac 12.g$.

In this case the condition $e=g(u,u)\neq 0$ is not used and $u$ could be a
non-null contravariant vector field ($e\neq 0$) as well as a null
contravariant vector field ($e=0$).

(c) One of the vector fields $b$ and $q$ or both vectors could also be
related to the Weyl's vector in a $(\overline{L}_n,g)$-space. If we
represent $\nabla _ug$ by means of its trace-free part and its trace part in
the form 
\begin{equation}  \label{2.21}
\nabla _ug=\,^s\nabla _ug+\frac 1n.Q_u.g\text{ , \thinspace \thinspace
\thinspace \thinspace \thinspace \thinspace \thinspace \thinspace \thinspace
\thinspace \thinspace \thinspace }\dim M=n\text{ ,}
\end{equation}

where $\overline{g}[^s\nabla _ug]=0$, $Q_u=\overline{g}[\nabla _ug]=g^{%
\overline{k}\overline{l}}.g_{kl;j}.u^j=Q_j.u^j$, \thinspace $Q_j=g^{%
\overline{k}\overline{l}}.g_{kl;j}$.

\textit{Remark}. The covariant vector $\overline{Q}=\frac 1n.Q=\frac
1n.Q_j.dx^j $ is called Weyl's vector. The operator $^s\nabla _u=\nabla
_u-\frac 1n.Q_u$ is called trace-free covariant operator.

(c$_1$) If we chose $b=\frac 1e.u$ and $q=\overline{g}(\overline{Q})=%
\widetilde{Q}$, then 
\begin{equation}  \label{2.22}
^c\nabla _u=\nabla _u-[\overline{g}(^F\omega )+g(\widetilde{Q},u).\overline{g%
}(g)-\frac 12.\overline{g}(\nabla _ug)]\text{ ,}
\end{equation}

and $^c\nabla _u\xi =0$ will have the form 
\begin{equation}  \label{2.23}
\nabla _u\xi =\overline{g}[(^F\omega )(\xi )]+g(\widetilde{Q},u).\xi -\frac
12.\overline{g}[(\nabla _ug)(\xi )]\text{ .}
\end{equation}

If the vector field $u$ is orthogonal to $\widetilde{Q}$, i. e. $g(%
\widetilde{Q},u)=0$, then $\nabla _u\xi =\overline{g}[(^F\omega )(\xi )%
]-\frac 12.\overline{g}[(\nabla _ug)(\xi )]$ and we have a generalized
Fermi-Walker transport of type $C$ along a non-null vector field $u$.

(c$_2$) If we chose $b=q=\overline{Q}$ in the expression for $^c\nabla _u$,
we will have the relation 
\begin{equation}  \label{2.24}
^c\nabla _u\xi =\nabla _u\xi -\{g(\widetilde{Q},u).[\overline{g}(^F\omega
)(\xi )+\xi ]-\frac 12.\overline{g}(\nabla _ug)(\xi )\}\text{ .}
\end{equation}

Then $^c\nabla _u\xi =0$ is equivalent to 
\begin{equation}  \label{2.25}
\nabla _u\xi =g(\widetilde{Q},u).[\overline{g}(^F\omega )(\xi )+\xi ]-\frac
12.\overline{g}(\nabla _ug)(\xi )\text{ . }
\end{equation}

For an orthogonal to $\widetilde{Q}$ vector field $u$ [$g(\widetilde{Q},u)=0$%
] we obtain a Fermi-Walker transport of type $B$ for the vector field $\xi
:\nabla _u\xi =-\frac 12.\overline{g}[(\nabla _ug)(\xi )]$ as in the case
(b) for $l=0$. Here $u$ could also be a non-null ($e\neq 0$) or null ($e=0$)
contravariant vector field.

\textit{Remark}. The assumption $g(\widetilde{Q},u)=0$ contradicts to the
physical interpretation of $\widetilde{Q}$ as a vector potential $A$ of the
electromagnetic field in $W_n$-spaces ($n=4$) \cite{Landau} because of the
relation $g(A,u)=e_0.[g(u,u)]/[g(R,u)\neq 0$. In our case here $\widetilde{Q}
$ could be more related to the Lorentz force $_LF=\overline{g}[F(u)]$ than
with $A$. $F$ is the electromagnetic tensor: $F=dA=(A_{j,i}-A_{i,j}).dx^i%
\wedge dx^j$.

\section{Conformal transports for null vector fields}

$l_\xi ^2=0$ is fulfilled for a null contravariant vector field $\xi $. Then
the condition for a conformal transport of $\xi $%
\begin{equation}  \label{3.1}
l_\xi .ul_\xi =\pm \frac 12.[(\nabla _ug)(\xi ,\xi )+2.g(\overline{A}_u\xi
,\xi )]=f(u).l_\xi ^2\text{ }
\end{equation}

with $\overline{A}_u=\overline{g}(C(u))=g(b,u).\overline{g}(^F\omega
)+g(q,u).\overline{g}(g)-\frac 12.\overline{g}(\nabla _ug)$ is fulfilled
identically for $\forall f(u)\in C^r(M)$. For two null contravariant vector
fields $\xi $ and $\eta $ [$l_\xi ^2=0$, $l_\eta ^2=0$] under a conformal
transport the relation 
\begin{equation*}
l_\xi .l_\eta .u[\cos (\xi ,\eta )]=(\nabla _ug)(\xi ,\eta )+g(\overline{A}%
_u\xi ,\eta )+g(\xi ,\overline{A}_u\eta )=0\text{ } 
\end{equation*}

is also identically fulfilled.

\textit{Remark}. Contravariant null vector fields fulfil also identically
the conditions for Fermi-Walker transports.

Therefore, \textit{every contravariant null vector field }$\xi $\textit{\
obeys automatically the conditions for a conformal transport. }Moreover,%
\textit{\ the (right) angle between two contravariant null vector fields }$%
\xi $\textit{\ and }$\eta $\textit{\ is automatically preserved under a
conformal transport.}

If the contravariant vector field $u$ is a null vector field [$e=g(u,u)=0$]
we cannot choose $b$ as $b=\frac 1e.u$ but we can consider it as $b=u$. In
such a case $g(b,u)=g(u,u)=0$ and $\overline{A}_u$ will have the form 
\begin{equation}  \label{3.2}
\overline{A}_u=g(q,u).\overline{g}(g)-\frac 12.\overline{g}(g)=\overline{g}%
(C(u))\text{ .}
\end{equation}

If we define further $q$ as a contravariant null vector field [$g(q,q)=0$],
then $q$ could be written in the form $q=\alpha .u$ [$\alpha \in C^r(M)$]
and $\overline{A}_u$ will have the form 
\begin{equation}  \label{3.3}
\overline{A}_u=-\frac 12.\overline{g}(\nabla _ug)\text{ }
\end{equation}

identical with the form of $\overline{A}_u$ for a Fermi-Walker transport of
type $B$.

\section{The length of a contravariant vector field as harmonic oscillator
over $(\overline{L}_n,g)$-spaces}

Let us now try to find out a solution of the following problem: under which
conditions the length $l_\xi $ of a non-null vector field $\xi $ moving
under conformal transport along a contravariant vector field $u$ could
fulfil the equation of a harmonic oscillator in the form 
\begin{equation}  \label{4.1}
\frac{d^2l_\xi }{d\tau ^2}+\omega _0^2.l_\xi =0\text{ , \thinspace
\thinspace \thinspace \thinspace \thinspace \thinspace \thinspace \thinspace
\thinspace \thinspace }\omega _0^2=\text{ const. }\geq 0\text{ , \thinspace
\thinspace }u^i=\frac{dx^i}{d\tau }\text{ , }\,\,\,\,\,\,u=\frac d{d\tau }%
\text{ . \thinspace \thinspace }
\end{equation}

Lets a contravariant non-null vector field $\xi $ be given, moving under
conformal transport along a trajectory $x^k(\tau )$ with the tangential
vector $u=\frac d{d\tau }$ in a $(\overline{L}_n,g)$-space. The change of
the length of $\xi $ along the trajectory is given as [see (\ref{2.1}) and (%
\ref{2.14})] 
\begin{equation}  \label{4.2}
\frac{dl_\xi }{d\tau }=g(q,u).l_\xi \text{ , \thinspace \thinspace
\thinspace \thinspace \thinspace \thinspace \thinspace \thinspace \thinspace
\thinspace \thinspace \thinspace \thinspace }\xi ,q,u\in T(M)\text{ .}
\end{equation}

Since $q$ is an arbitrary given contravariant vector field we can specify
its structure in a way allowing us to consider $g(q,u)$ as an invariant
scalar function depending only of the parameter $\tau $ of the trajectory $%
x^k(\tau )$. A possible choice of $q$ fulfilling this precondition is 
\begin{equation}  \label{4.3}
q=\omega (\tau ).\frac 1e.u\text{ , \thinspace \thinspace \thinspace
\thinspace \thinspace \thinspace \thinspace \thinspace \thinspace \thinspace
\thinspace }\omega (\tau )=\omega (x^k(\tau ))\in C^r(M)\text{ , \thinspace
\thinspace \thinspace \thinspace }r\geq 1\text{, \thinspace \thinspace
\thinspace \thinspace }e=g(u,u)\neq 0\text{ . }
\end{equation}

Then we have 
\begin{equation}  \label{4.4}
\frac{dl_\xi }{d\tau }=\omega (\tau ).l_\xi
\end{equation}

\textit{Remark}. The solution of this equation for $l_\xi $ can be found in
the form 
\begin{equation}  \label{4.5}
l_\xi =l_{\xi 0}.\exp (\int \omega (\tau ).d\tau )\text{ , \thinspace
\thinspace \thinspace \thinspace \thinspace \thinspace \thinspace \thinspace
\thinspace }l_{\xi 0}=\text{ const., \thinspace \thinspace \thinspace }l_\xi
>0\text{ ,}
\end{equation}

where $d(\log l_\xi )=\omega (\tau ).d\tau $, \thinspace \thinspace $\log
l_\xi =\int \omega (\tau ).d\tau +C_0$, $C_0=$ const.

After differentiation of the expression (\ref{4.4}) along $\tau $ and after
further use of the same expression we obtain 
\begin{equation*}
\frac{d^2l_\xi }{d\tau ^2}=\frac{d\omega (\tau )}{d\tau }.l_\xi +\omega
(\tau ).\frac{dl_\xi }{d\tau }=\frac{d\omega (\tau )}{d\tau }.l_\xi +[\omega
(\tau )]^2.l_\xi = 
\end{equation*}
\begin{equation}  \label{4.6}
=\left\{ \frac{d\omega (\tau )}{d\tau }.+[\omega (\tau )]^2\right\} .l_\xi 
\text{ .}
\end{equation}

The equation for $d^2l_\xi /d\tau ^2$ could be written in the form 
\begin{equation}  \label{4.7}
\frac{d^2l_\xi }{d\tau ^2}-\left\{ \frac{d\omega (\tau )}{d\tau }.+[\omega
(\tau )]^2\right\} .l_\xi =0\text{ .}
\end{equation}

It could have the form of an oscillator equation if the arbitrary given
until now function $w(\tau )$ fulfils the equation 
\begin{equation}  \label{4.8}
\frac{d\omega (\tau )}{d\tau }.+[\omega (\tau )]^2=-\,\,\omega _0^2\leq 0%
\text{ }
\end{equation}

and could be determined by means of this (additional) condition. The last
equation is a Riccati equation \cite{Kamke}. By the use of the substitution $%
u^{\prime }=\omega .u$, $u^{\prime }=du/d\tau $, it could be written in the
form of a homogeneous harmonic oscillator equation for $u$%
\begin{equation}  \label{4.9}
u^{\prime \prime }+\omega _0^2.u=0\text{ ,}
\end{equation}

where 
\begin{equation*}
\omega (\tau )=\frac{u^{\prime }(\tau )}{u(\tau )}\text{ , \thinspace
\thinspace \thinspace \thinspace \thinspace \thinspace \thinspace }\omega
^{\prime }=-\frac{(u^{\prime })^2}{u^2}+\frac{u^{\prime \prime }}u\text{ ,
\thinspace \thinspace \thinspace \thinspace \thinspace }\omega ^{\prime
}+\omega ^2=-\omega _0^2\text{ .} 
\end{equation*}

For $\omega _0^2\geq 0$ (\ref{4.9}) has the solutions:

(a) $\omega _0^2=0:u^{\prime }=C_2.\tau $ , \thinspace \thinspace \thinspace 
$u=C_1+C_2.\tau $ , $C_1,C_2=$ const.

(b) $\omega _0^2>0:u=C_1.\cos \omega _0.\tau +C_2.\sin \omega _0.\tau $ , $%
C_1,C_2=$ const.

The solution for $\omega $ will have the forms:

(a) $\omega _0^2=0$: 
\begin{equation}  \label{4.10}
\omega (\tau )=\frac{C_2.\tau }{C_1+C_2.\tau }=\frac \tau {\frac{C_1}{C_2}%
+\tau }=\frac \tau {a+\tau }\text{ , \thinspace \thinspace \thinspace
\thinspace \thinspace \thinspace }a=\frac{C_1}{C_2}=\text{const., \thinspace
\thinspace \thinspace \thinspace \thinspace }C_2\neq 0\text{ .}
\end{equation}

(b) $\omega _0^2>0:u=C_1.\cos \omega _0.\tau +C_2.\sin \omega _0.\tau $,
\thinspace $u^{\prime }=\omega _0.[C_2.\cos \omega _0.\tau -C_1.\sin \omega
_0.\tau ]$, $C_1$, $C_2=$ const., 
\begin{equation}  \label{4.11}
\omega (\tau )=\frac{u^{\prime }}u=\omega _0.\frac{C_2.\cos \omega _0.\tau
-C_1.\sin \omega _0.\tau }{C_1.\cos \omega _0.\tau +C_2.\sin \omega _0.\tau }%
=\omega _0.\frac{1-\frac{C_1}{C_2}.\tan \omega _0.\tau }{\frac{C_1}{C_2}%
+\tan \omega _0.\tau }\text{ , \thinspace \thinspace \thinspace }C_2\neq 0%
\text{ ,}
\end{equation}
\begin{equation}  \label{4.12}
\omega (\tau )=\omega _0.\frac{1-a.\tan \omega _0.\tau }{a+\tan \omega
_0.\tau }\text{ , \thinspace \thinspace \thinspace \thinspace \thinspace
\thinspace \thinspace \thinspace }a=\frac{C_1}{C_2}=\text{ const.}
\end{equation}

(b$_1$) For $C_2=0:\omega (\tau )=-\omega _0.\tan \omega _0.\tau $ .

(b$_2$) For $C_1=0:\omega (\tau )=\omega _0.\cot \omega _0.\tau =\omega
_0/\tan \omega _0.\tau $.

Therefore, if $l_\xi $ should be a harmonic oscillator in a $(\overline{L}%
_n,g)$-space with a given constant frequency $\omega _0$ the function $%
\omega (\tau )$ has to obey a Riccati equation determining its structure. In
this case, the length $l_\xi $ will be a homogeneous harmonic oscillator
obeying at the same time the equation 
\begin{equation}  \label{4.13}
\frac{dl_\xi }{d\tau }=\omega _0.\frac{1-a.\tan \omega _0.\tau }{a+\tan
\omega _0.\tau }.l_\xi \text{ , \thinspace \thinspace \thinspace \thinspace
\thinspace \thinspace \thinspace \thinspace \thinspace \thinspace \thinspace
\thinspace \thinspace \thinspace }a=\frac{C_1}{C_2}=\text{ const.}\neq 0%
\text{ ,}
\end{equation}

which appears in general as a solution of the homogeneous harmonic
oscillator's equation 
\begin{equation*}
\frac{d^2l_\xi }{d\tau ^2}+\omega _0^2.l_\xi =0\text{ , \thinspace
\thinspace \thinspace \thinspace \thinspace \thinspace \thinspace \thinspace
\thinspace \thinspace }\omega _0^2=\text{ const. }\geq 0\text{ .} 
\end{equation*}

If we could simulate by the use of an appropriate experimental device the
change $(dl_\xi /d\tau )$ as given in its equation (\ref{4.13}), then we can
be sure that $l_\xi $ will swing as a harmonic oscillator with the frequency 
$\omega _0$ over a $(\overline{L}_n.g)$-space. This could allow us to
investigate experimentally the influence of physical interactions on the
length $l_\xi $ of a vector field $\xi $ moving under a conformal transport
over a $(\overline{L}_n,g)$-space. On the other hand, if we can register (as
an observer) from our basic trajectory a change of $l_\xi $ in accordance
with the equation for $(dl_\xi /d\tau )$, then we can conclude that $l_\xi $
moves as harmonic oscillator under the external (or internal) forces. Since
the considered here problem is related to the kinetic aspect of the motion
of $l_\xi $, the dynamic aspect should be introduced by means of a concrete
field (gravitational) theory, which is not a subject of the above
considerations.

\subsection{Geometrical and physical interpretation of the function $\protect%
\omega (\protect\tau )$}

The covariant derivative $\nabla _u\xi $ of a contravariant vector field $%
\xi $ along a contravariant vector field $u$ in a $(\overline{L}_n,g)$-space
can be represented in the form \cite{Manoff-5} 
\begin{equation}  \label{4.14}
\nabla _u\xi =\frac{\overline{l}}e.u+\overline{g}[h_u(\nabla _u\xi )]=\frac{%
\overline{l}}e.u+\,_{rel}v\text{ , \thinspace \thinspace \thinspace
\thinspace \thinspace \thinspace \thinspace \thinspace \thinspace \thinspace
\thinspace \thinspace \thinspace }\overline{l}=g(\nabla _u\xi ,u)
\end{equation}

where 
\begin{equation}  \label{4.15}
_{rel}v=\overline{g}[h_u(\nabla _u\xi )]=\overline{g}(h_u)(\frac
le.a-\pounds _\xi u)+\overline{g}[d(\xi )]\text{ , \thinspace \thinspace
\thinspace }h_u=g-\frac 1e.g(u)\otimes g(u)\text{ ,}
\end{equation}
\begin{equation}  \label{4.16}
a=\nabla _uu\text{ , \thinspace \thinspace \thinspace \thinspace \thinspace
\thinspace \thinspace \thinspace }d=\sigma +\omega +\frac 1{n-1}.\theta .h_u%
\text{ , \thinspace \thinspace }g(\nabla _u\xi ,u)=ul-(\nabla _ug)(\xi
,u)-g(\xi ,a)\text{ .}
\end{equation}

$\sigma $ is the shear velocity tensor (shear), $\omega $ is the rotation
velocity tensor (rotation), $\theta $ is the expansion velocity invariant
(expansion), $d$ is the deformation velocity tensor (deformation), $l=g(\xi
,u)$, $_{rel}v$ is the relative velocity. If $\xi $ is an orthogonal to $u$
vector field [$\xi =\xi _{\perp }=\overline{g}[h_u(\xi )]$], then $l=0$ and
under the additional precondition $\pounds _\xi u=-\pounds _u\xi =0$ the
expression for $_{rel}v$ will take the form 
\begin{equation}  \label{4.17}
_{rel}v=\overline{g}[d(\xi _{\perp })]=\overline{g}[\sigma (\xi _{\perp })]+%
\overline{g}[\omega (\xi _{\perp })]+\frac 1{n-1}.\theta .\xi _{\perp }\text{
.}
\end{equation}

The rate of change of the length $l_{\xi _{\perp }}$ of the vector field $%
\xi _{\perp }$ (along the vector field $u=\frac d{d\tau }$) in $\overline{U}%
_n$- or $\overline{V}_n$-spaces [$\nabla _ug=0$ for $\forall u\in T(M)$],
under the conditions $l=0$ [$\xi =\xi _{\perp }=\overline{g}[h_u(\xi )]$]
and $\pounds _u\xi =0$, can be found in the form 
\begin{equation}  \label{4.18}
ul_{\xi _{\perp }}=\frac{dl_{\xi _{\perp }}}{d\tau }=\pm \frac 1{l_{\xi
_{\perp }}}.d(\xi _{\perp },\xi _{\perp })=\pm \frac 1{l_{\xi _{\perp
}}}.\sigma (\xi _{\perp },\xi _{\perp })+\frac 1{n-1}.\theta .l_{\xi _{\perp
}}\text{ ,\thinspace \thinspace \thinspace \thinspace \thinspace \thinspace
\thinspace \thinspace \thinspace \thinspace \thinspace \thinspace \thinspace
\thinspace }l_{\xi _{\perp }}\neq 0\text{ .}
\end{equation}

\textit{Remark}. The sign $\pm $ depends on the sign of the metric $g$ (for $%
n=4 $, sign $g=\pm 2$).

If a $U_n$- or a $V_n$-space admits a shear-free non-null auto-parallel
vector field $u$ ($\nabla _uu=a=0$), then $\sigma =0$ and 
\begin{equation}  \label{4.19}
\nabla _u\xi _{\perp }=\,_{rel}v=\overline{g}[\omega (\xi _{\perp })]+\frac
1{n-1}.\theta .\xi _{\perp }=\overline{g}[\omega (\xi _{\perp })]+\frac
1{n-1}.\theta .\overline{g}[g(\xi _{\perp })]=
\end{equation}

\begin{equation}  \label{4.20}
=\overline{g}[C(u)(\xi _{\perp })]=\overline{A}_u\xi _{\perp }\text{
,\thinspace \thinspace \thinspace \thinspace \thinspace \thinspace
\thinspace \thinspace \thinspace \thinspace \thinspace \thinspace }%
C(u)=\omega +\frac 1{n-1}.\theta .g\text{ .}
\end{equation}
\begin{equation}  \label{4.21}
\frac{dl_{\xi _{\perp }}}{d\tau }=\frac 1{n-1}.\theta .l_{\xi _{\perp }}%
\text{ .}
\end{equation}

Therefore, the vector field $\xi _{\perp }$ undergoes a conformal transport
along $u$. A comparison with (\ref{4.4}) show us that in this case we can
choose the arbitrary given function $\omega (\tau )$ as 
\begin{equation}  \label{4.22}
\omega (\tau )=\frac 1{n-1}.\theta \text{ .}
\end{equation}

The last fact leads to the conclusion that $\omega (\tau )$ could be related
to the expansion velocity $\theta $ in a $\overline{U}_n$- or a $\overline{V}%
_n$-space. In $(\overline{L}_n,g)$-spaces it could preserve its
interpretation.

\section{Conclusions}

In the present paper we have considered types of transports more general
than the Fermi-Walker transports. They are called conformal transports over $%
(\overline{L}_n,g)$-spaces. In an analogous way as in the case of
Fermi-Walker transports a conformal covariant differential operator and its
corresponding conformal derivative are determined and discussed over $(%
\overline{L}_n,g)$-spaces. Different special types of conformal transports
are considered inducing also Fermi-Walker transports for orthogonal vector
fields as special cases. Conditions under which the length of a non-null
contravariant vector field will swing as a homogeneous harmonic oscillator
with given frequency are established. The results obtained regardless of any
concrete field (gravitational) theory could have direct applications in such
types of theories.

\begin{center}
\textsc{Acknowledgments}
\end{center}

The author is grateful to Prof. F.-W. Hehl, Prof. St. Dimiev and Prof. K.
Sekigawa for their friendly help and support in preparing this report. This
work is also supported in part by the National Science Foundation of
Bulgaria under Grant No. F-642.

\end{document}